\documentclass[twocolumn,superscriptaddress,amsmath,amssymb,floatfix,aps,notitlepage,nokeys,prl,longbibliography,nofootinbib]{revtex4-1}

\usepackage{titlesec}  % Package to customize titles
\usepackage{textcase}  % For capitalizing words

\titleformat{\section} % Redefine the section style
  {\raggedright\bfseries} % Left-aligned and bold
  {\thesection}{1em}{}   % Keeps the section number
 \usepackage{epstopdf}
\usepackage{fancyhdr} % Package to control page headers and footers

\usepackage{geometry}
\usepackage{csquotes}
\usepackage{mathrsfs, xr-hyper}
\usepackage{physics}
\usepackage{hyperref, amssymb, amsbsy, amsmath, latexsym, dsfont, array, layout, graphics,mathrsfs,braket,amsfonts,amsthm,
  amssymb,graphicx,subfigure, dcolumn,youngtab,color,bm,mathtools,braket,verbatim,url,cleveref,natbib,hypernat,extarrows,inputenc,multirow,bbm,dsfont,soul,ulem}

\usepackage{commath}
\usepackage[mathlines]{lineno}
\usepackage{tikz}
\usepackage{accents}

\begin{document}
\title{Lattice Materials with Topological States Optimized On-Demand}

\author{Pegah Azizi}
\altaffiliation{These authors contributed equally to this work}
\affiliation{%
Department of Civil, Environmental, and Geo- Engineering,
University of Minnesota, Minneapolis, MN 55455, US}
\author{Rahul Dev Kundu}
\altaffiliation{These authors contributed equally to this work}
\affiliation{%
Department of Civil and Environmental Engineering, University of Illinois at Urbana-Champaign, Urbana, IL, 61801, USA
}
\author{Weichen Li}
\affiliation{%
Department of Civil and Environmental Engineering, University of Illinois at Urbana-Champaign, Urbana, IL, 61801, USA
}
\author{Kai Sun}
\affiliation{%
Department of Physics, University of Michigan, Ann Arbor, MI 48109, USA
}
\author{Xiaojia Shelly Zhang}
\email{zhangxs@illinois.edu}
\affiliation{%
Department of Civil and Environmental Engineering, University of Illinois at Urbana-Champaign, Urbana, IL, 61801, USA
}
\author{ Stefano Gonella}%
\email{sgonella@umn.edu}
\affiliation{%
Department of Civil, Environmental, and Geo- Engineering,
University of Minnesota, Minneapolis, MN 55455, US}%

\begin{abstract}

\textbf{Abstract:} 
Topological states of matter, first discovered in quantum systems, have opened new avenues for wave manipulation beyond the quantum realm. In elastic media, realizing these topological effects requires identifying lattices that support the corresponding topological bands. However, among the vast number of theoretically predicted topological states, only a small fraction has been physically realized.
To close this gap, we present a new strategy capable of systematically and efficiently discovering metamaterials with any desired topological state. Our approach builds on topological quantum chemistry (TQC), which systematically classifies topological states by analyzing symmetry properties at selected wavevectors. Because this method condenses the topological character into mathematical information at a small set of wavevectors, it encodes a clear and computationally efficient objective for topology optimization algorithms. We demonstrate that, for certain lattice symmetries, this classification can be further reduced to intuitive morphological features of the phonon band structure. By incorporating these band morphology constraints into topology optimization algorithms and further fabricating obtained designs, we enable the automated discovery and physical realization of metamaterials with targeted topological properties. This methodology establishes a new paradigm for engineering topological elastic lattices on demand, addressing the bottleneck in material realization and paving the way for a comprehensive database of topological metamaterial configurations.\\

\textbf{Significance statement:} % (120 words)}:
The legacy of decades of phononics research is a sharp awareness of the wave manipulation capabilities of elastic metamaterials. The incorporation of optimization methods is also bringing metamaterials technology closer to widespread adoption in engineering design. The most recent leap broadening the horizon of metamaterials functionalities is traceable to the injection of notions of topological states of matter. However, these physics have yet to be systematically incorporated into design frameworks, due to lacking efforts effectively encoding principles of topology into drivers of optimization. Offering a much-needed synthesis between these perspectives, we provide criteria to distill salient topological attributes and recast them as objectives of optimization algorithms, paving the way for computerized design and physical realization of libraries of metamaterials with on-demand topological states.\\

\textbf{Keywords:} topology optimization $|$ topological mechanics  $|$ mechanical metamaterials $|$ experimental demonstration $|$ wave control % At least three keywords are required at submission. Please provide three to five keywords, separated by the pipe symbol.
\end{abstract}

\maketitle
The ability to control mechanical waves and vibrations is a highly sought-after attribute in material systems across engineering and materials science applications. Homogeneous solid media lack the architectural complexity required to manipulate the propagation of elastic waves beyond conventional regimes. This limitation has sparked growing interest in \textit{architected materials} and \textit{ metamaterials}. Metamaterials owe their properties not to their composition but to the geometry of their internal architecture which, in the case of \textit{lattice materials}, is spatially periodic. Through careful design, elastic metamaterials provide unprecedented control of wave propagation, enabling notable phenomena such as bandgap (BG) opening and tuning, directivity, negative refraction and superfocusing~\cite{Li2017BGap_APL,Fang2018Bgap_JAppMech,SUGINO2018Bgap_JMPS,Bastiaan2014_PRL_metamat,Zhu2014_metamat,Bertoldi2017_natrev_metamat,YU2018_progmatsci_metamat,Surjadi2019_metamat_advengmat,Stenseng_metamat_2024advsci}.

An additional dimension to this array of capabilities has emerged from the injection of notions of \textit{topology}. Through a topological classification of wave descriptors defined in the frequency-wavevector space (\textit{k-space topology}), and by adapting to elastodynamics selected concepts of topological phases of matter originally introduced to describe quantum and electronic phenomena, we can elicit and interpret effects that elude conventional phononics analysis. This philosophy has opened the field of \textit{topological mechanical metamaterials}, material systems that exhibit static and dynamic regimes that are robust against defects and perturbations, e.g., polarized elasticity and one-way edge and interface states~\cite{Kane_Lubensky_Nphys_2014,Rocklin_et_all_natcom_2017,Raj_Ruzzene_QVHE_NewJournalPhys_2017,Vila_Pal_Ruzzene_PRB_VALLEY,Mou_etall_Valley_NatMat_2018,Yu_etall_topopseudo_NatCom_2018,Mousavi_etall_topoprotected_NatCom_2015,Xiao_acoustictopo_NatPhys_2015,He_acousTI_NatPhys_2016,Nash_etall_gyroscop_pnas_2015}. These topologically protected phenomena are associated with the presence of topological bands in the phonon spectrum. Consequently, the ability to recognize the existence of such bands—and, more importantly, to promote their emergence when designing new materials—has become a crucial objective. 

\begin{figure*}[t]
\includegraphics[width=1.0\textwidth]{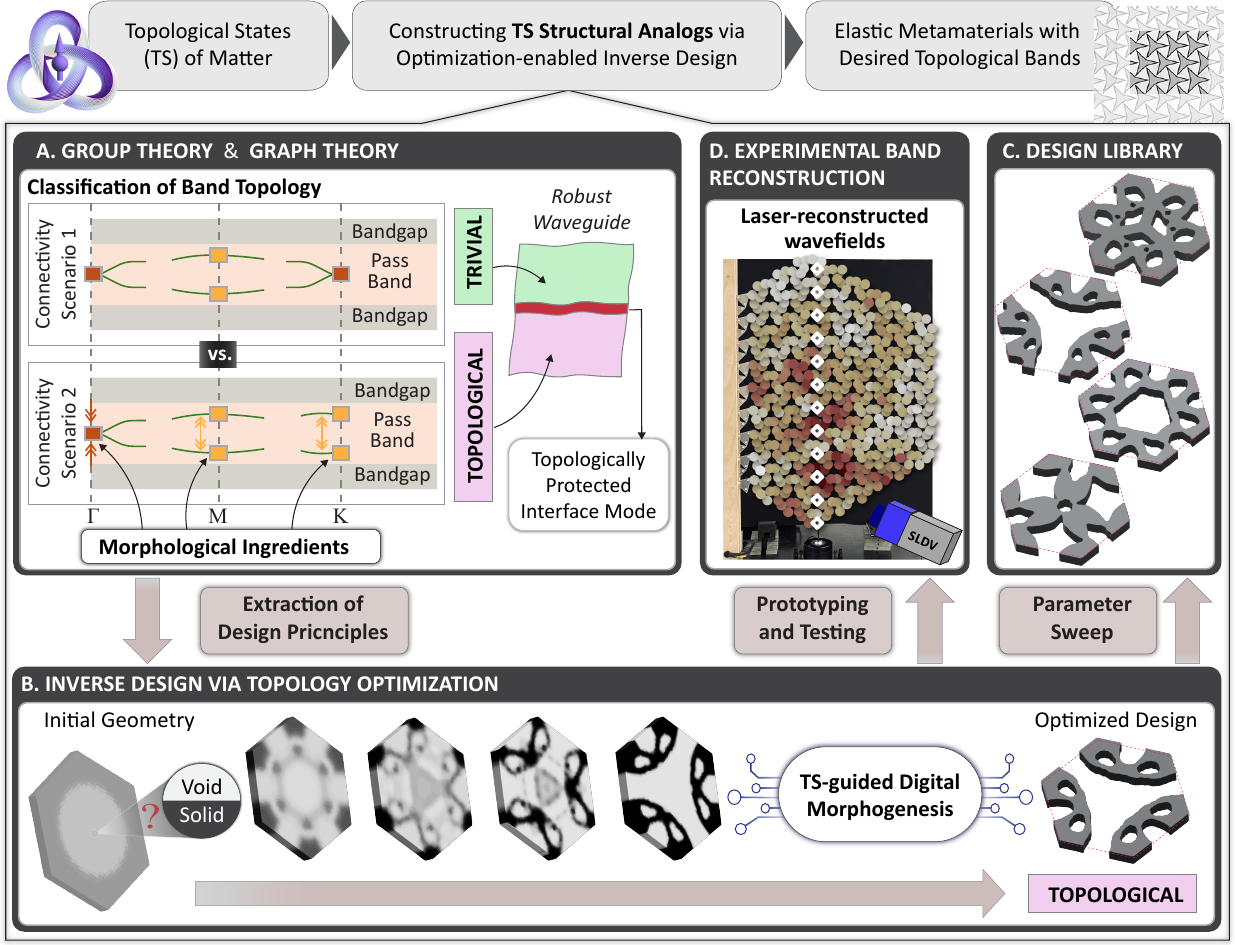}
\caption{\label{fig:FramWrk}Conceptual schematic of the proposed framework. (A) Two connectivity scenarios within the $p31m$ symmetry group, corresponding to trivial and topological bands as identified by TQC-enabled band classification, along with a potential application. Morphological band features identified as key drivers for the topolgy optimization (TO) algorithm. (B) Digital morphogenesis inverse design process via the proposed TO framework, enabling automatic discovery of metamaterials with any desired topological features, showing intermediate steps and an optimized configuration. (C) A selection of generated configurations, demonstrating the diverse patterning achieved. (D) Laser vibrometry testing of a fabricated prototype, used to experimentally reconstruct and validate the band diagram.}
\end{figure*}

The recently-developed framework of \textit{Topological Quantum Chemistry} (TQC) offers a streamlined approach to classifying and identifying topological bands % materials
based on symmetry arguments~\cite{bradlyn_et_all_TQC_Nat_2017,vergniory_et_all_Billcrys_Phys.Rev.E_2017,cano_et_all_TQC_Phys.Rev.B_2018,kruthoff2017topological,po2017symmetry,Po_2020_SYMindic,khalaf2018symmetry,Bradlyn_et_all_fragile_PhysRevB_2019,Song_et_all_fragile_science_2020}. By relying exclusively on the inspection and classification of irreducible representation labels (the so-called \textit{irreps}) evaluated at the high-symmetry points (HSPs) of the Brillouin zone (BZ), TQC simplifies the identification of symmetry-protected topological states (TSs), offering an efficient and practical alternative to computationally intensive methods that traditionally require detailed knowledge of Bloch wave functions across the entire BZ to define topological invariants. In essence, the irreps capture how the eigenfunctions—in elastodynamics, the unit cell mode shapes—transform under various symmetry operators. Formally, a band structure is deemed topologically nontrivial if its irreps cannot be matched to those of any atomic insulator, indicating a state that, by definition, cannot be described by %admit 
symmetric localized Wannier functions~\cite{bradlyn_et_all_TQC_Nat_2017}. The Bilbao Crystallographic Server (BCS) compiles all possible atomic insulators by combining lattice and orbital types in terms of irreps, thus providing a comprehensive reference to determine whether a given band structure is topologically trivial or hosts symmetry-protected topological states~\cite{aroyo_et_all_Billcrys_BulgChemCommun_2011,Aroyo_et_all_Billcrys_2006I,Aroyo_et_all_Billcrys_2006II,Elcoro2017_bilbao}. Recently, researchers have invoked the TQC framework to identify topological bands in photonic crystals~\cite{Paz_PhysRevResearch_2019_photonicsTQC}, acoustic~\cite{peri_science_2020_acousticTQC}, and phononic metamaterials~\cite{Siddhartha_thesis,Azizi-et-al_Fragile-kagome_PRL_2023,Zhang_PhysRevApplied_2023_elasticTQC,Azizi_2024_DW_PhysRevB}.  

A promising approach for discovering mechanical metamaterials with prescribed properties is inverse design through topology optimization (TO), a design methodology based on strategic spatial allocation of material phases in a given design domain guided by mechanics-based algorithms~\cite{Bendsoe1988,Bendsoe2004_TObook,Wang2021}. Providing broader design freedom than intuition- or trial-and-error-based methods, TO has been applied to the inverse design of phononic structures with the goal of optimizing certain wave control capabilities, such as BG widening or lowering~\cite{Sigund-Jensen_TO-Bandgap_PRSA_2003, Dalklint2022_intro,Liu2024_intro, Dong2024_intro,Wang2019_ultrawideBG,Luo2022_soft,Wu2023_prescribedBG,Vatanabe2014_piezoBG,Jia2024_BG}, negative refraction~\cite{Christiansen2016_negativerefrac}, acoustic and photonic cavities~\cite{Christiansen2015_acousticcavity,Wang2018_photoniccavity}, and elastic wave barrier and absorber~\cite{Vanhoorickx2016_wavebarrier,Matsushima2020_intro}. More recently, TO has been employed to design topological insulators by %incorporating various drivers, such as leveraging 
enforcing global signatures of topological behavior (e.g., the emergence of interface states) as optimization objectives~\cite{Christiansen2019_intro}, by promoting Dirac cones to obtain quantum spin Hall analogs~\cite{Nanthakumar2019_intro,Chen2021_intro,Lu2021_intro,Christiansen2019_spinHall} or quantum valley Hall analogs~\cite{Zhuang2022_intro,Zhang2022_intro}, or by directly optimizing topological invariants~\cite{Luo2023_intro}. Despite these advances, the direct encoding of phonon band topology criteria into TO frameworks remains largely unexplored due to the inherent challenges of mathematically capturing the unique conditions that underlie topological phenomena and translating them into efficient drivers for the optimization algorithms. In addition, theoretical studies have vastly outpaced experimental realizations of TSs, with only a relatively small subset being directly confirmed experimentally. 

This gap underscores the need for a versatile inverse design framework that systematically encodes the topological descriptors directly into the optimization process, thus enabling the discovery of metamaterials that display desired topological properties while retaining the structural characteristics that make them amenable to physical fabrication and testing. In parallel to the primary objective of guaranteeing the emergence of desired topological bands on-demand, such optimization framework could ideally be set up to simultaneously pursue other (secondary) desired targets, such as tuning BG onsets and widths to target desired operational frequency regimes. This set of goals involves the nontrivial task of distilling the topological requirements of bands into a parsimonious set of rigorous band descriptors that can be expressed mathematically and effectively incorporated into optimization algorithms. The TQC approach discussed above, which simplifies the topology of bands to mere symmetry considerations, constitutes an ideal tool around which we can construct our design strategy. 
%, thereby making the observation of topological phenomena more accessible in practice. 

In the remainder of this paper, we will propose a strategy to address these design and physical demonstration needs, leveraging the integration of two powerful and complementary tools: 1) a method for topological classification of bands, based on TQC and boosted by recent discoveries that link TQC criteria to band morphology requirements; 2) a TO strategy specifically tailored to promote the emergence of topological features through the incorporation of key band morphology descriptors as drivers of the optimization algorithms. The discovered metamaterial is then physically fabricated and experimentally demonstrated to possess the target topological properties. The conceptual steps of this strategy are summarized by pictorial highlights in Fig.~\ref{fig:FramWrk}. (A) We start from a general classification philosophy of band topology, rooted in principles of group and graph theory and formalized by TQC. Here, we invoke a key finding from our recent study, which pinpoints the emergence of band topology to few band morphological attributes that can be assessed through agile band inspection. (B) We encode these morphological requirements mathematically as objectives and constraints of TO algorithms to enable the automatic discovery of metamaterials. %in order to generate optimized designs with desired topological properties. 
(C) We comprehensively explore the non-convex design space of lattices to compile libraries of configurations with a common desired topological character as primary attribute and a plethora of secondary phononic attributes. (D) Finally, we test the dynamics of selected configurations using physical prototypes to confirm the emergence of the desired band topology through an a-posteriori assessment of the morphology of experimentally reconstructed bands. In the next section, we will discuss all these steps in detail.
\section{Results and Discussion}
\subsection{Fundamental Relations Between TQC and Band Morphology.} 
In this section, we outline the key conceptual steps of our proposed framework for designing lattice materials with desired TSs based on band morphology criteria. Our starting point leverages a key intuition gained in our previous study on the dynamics of structural kagome metamaterials (Azizi et al.~\cite{Azizi-et-al_Fragile-kagome_PRL_2023}). This study adopted the symmetry arguments of TQC to assess the topological character of bands, thus extending to elastic lattice materials a framework originally proposed for -and tested against- quantum and electronic systems. Recall that a key insight from TQC is the notion that the topological character of bands can be assessed by looking at certain symmetry indicators, known as ``irreps", which capture how the eigenfunctions transform under a number of symmetry operators. The landscapes of irreps for a band, or cluster of isolated bands, that are associated with trivial bands are cataloged in the BCS; hence, a band of a lattice belonging to a given symmetry class is deemed topological if it does not conform to any of the tabulated landscapes of irreps for that symmetry class.

\begin{figure*}[t!]
\includegraphics[width=1\textwidth]{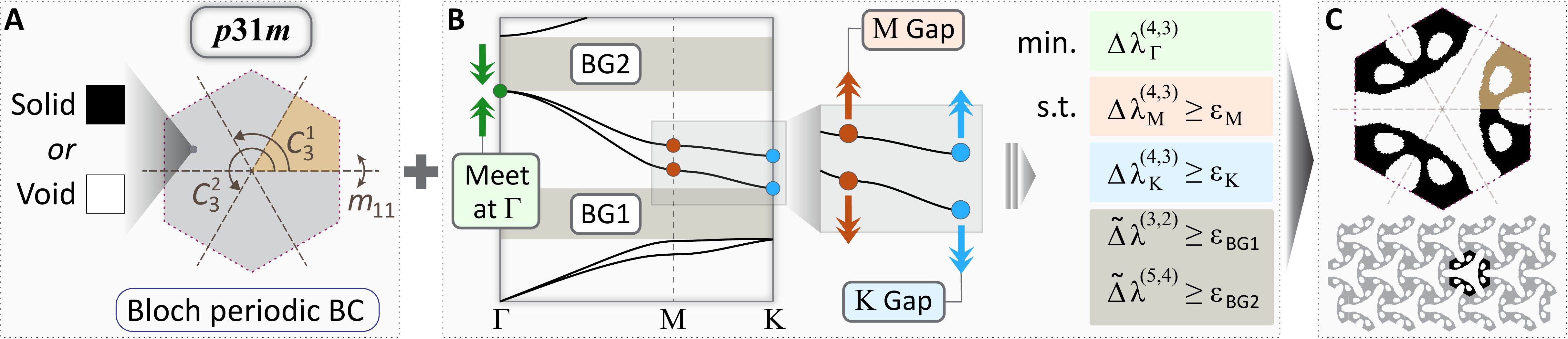}
\caption{\label{fig:TopOpt}Digital morphogenesis inverse design via band morphology-driven TO. (A) Design domain, boundary conditions (BC), and candidate material phases. (B) Target band morphology translated and encoded into an optimization formulation. (C) Discovered metamaterial design with desired topological features and its tessellated assembly.}
\end{figure*}

Furthermore, in~\cite{Azizi-et-al_Fragile-kagome_PRL_2023} a connection was demonstrated between the topological character of the bands and certain morphological attributes of the corresponding branches in the band diagram. It was shown that, for the considered symmetry class ($p31m$ wallpaper group), the existence of two isolated bands that cross exclusively at the HSP $\Gamma$, without retouching at any other HSPs, i.e., K and M, constitutes a necessary and sufficient condition to guarantee the topological character of those bands. In essence, the matter of topological classification %was shown to 
reduces to an agile and intuitive inspection of a parsimonious set of morphological branch descriptors, which serve as proxies for the TQC irreps. This powerful observation implies that one can ascertain the emergence of topology from experimental or numerical data by simply reconstructing the bands from the measured response and assessing their morphology - a routine operation in phononics that can be carried out agnostically, even in the absence of a precise model of the lattice cell. 
%In the specific case of structural kagome lattices, it was shown that, under certain geometric requirements of the inter-cell ligament-like connections, the lattices featured two isolated optical bands %(Fig.~\ref{fig:FramWrk}.B) 
%that were deemed topological.
%\sss{The study also tested the robustness of this topology by monitoring the evolution of the irreps when the bands are analyzed together with their neighbors: specifically, it was shown that adding the acoustic bands modified the irreps landscape in a way that trivialized the topology. For this reason, the topology of the mid-frequency bands was classified as \textit{fragile} (see SM, sec for details).} 
The idea, while conceived in the context of lattices with $p31m$ wallpaper group, is extendable in principle to a broader class of lattices, although other case-specific %band morphology 
attributes may need to be inferred for each case. 

% We now intend to show that
%\sss{The important link established between band topological character and band morphology} 
The ability to classify the mechanical performance in terms of a discrete set of parameters provides a golden platform for the development of optimization frameworks for structural and material design. %has even broader implications in the context of structural and material design. 
Not only can the morphology criteria be used \textit{directly} to characterize the topological character of the bands; they can also serve as drivers of an \textit{automated inverse design process} aimed at discovering new lattice configurations of arbitrary complexity with desired topological attributes. The idea is to construct an \textit{optimization algorithm} in which such criteria are embedded either as \textit{objective functions} or \textit{constraints} of the optimization problem. In the next section, we elucidate this idea targeting, as a benchmark example, the band morphology observed for finite-frequency topological bands in lattices of the $p31m$ wallpaper group.% (Fig.~\ref{Idea_1}(d-e)). 

\subsection{A Band Morphology-driven \\TO Strategy to Find TSs.}
%This special landscape of irreps is uniquely related to a distinct morphology of the bands and specifically to two ingredients 1) we have two isolated bands sandwiched between two bandgaps; 2) these bands cross at Γ without touching at any other HSP. We can show that,
%for our lattice symmetry class, these morphological features are sufficient and necessary ingredients for (fragile) topology. This powerful observation implies that we can validate the emergence of fragile topology from experimental or numerical data by simply reconstructing the bands from the measured response and assessing their morphology, an inference that can be carried out agnostically, even when we lack a precise model of the lattice cell. Not only can the morphology criteria be used to assess the topological character of the bands, they can also be used as drivers of an automated design process geared to identify lattice configurations of arbitrary complexity with desired topological attributes.
%This subsection describes the band morphology-driven topology optimization (TO) strategy that enables automated discovery of lattice metamaterials with unconventional phononic behaviors. We use the $p31m$ wallpaper group lattices as an example, and seek To obtain lattice metamaterials exhibiting fragile topological state. 
The first step toward automated discovery of lattice configurations % metamaterials 
via band morphology-driven TO is to identify the primary design principles, i.e., the desired topological attributes of the targeted phononic behavior. In the case of the $p31m$ wallpaper group lattice class, the primary design principles for achieving topological bands distill to the realization of the following set of band characteristics:
\begin{itemize}
\item enforce a band crossing at the HSP $\Gamma$, while maintaining gaps at the HSPs M and K between bands 3 and 4;
\vspace{-0.1 in}
\item ensure the complete isolation of band 3 (4) from band 2 (5);
%\vspace{-0.1 in}
%\item complete isolation of band 4 from band 5.
%\vspace{-0.1 in}
\end{itemize}
The next step is to translate these requirements into mathematical expressions, i.e., differentiable functions that can be encoded in gradient-based optimization algorithms. The optimization tasks %desired band characteristics 
can be classified into two major categories -- a) merging or opening gaps between specific bands at HSPs only, and b) complete isolation of different bands. For the first category, we simply express the gaps between bands $m$ and $n$ at the HSP $t \in \{\Gamma, \text{K}, \text{M}\}$ %along the contour of the BZ 
as 
\begin{equation} \label{eq:gap_function}
    \Delta \lambda^{(m,n)}_t = \lambda_t^{(m)} - \lambda_t^{(n)}, \ \ (m \geq n),
\end{equation} 
where $\lambda_t$ represents the squared eigenfrequency $\omega_t$ obtained by solving the generalized eigenvalue problem $\left(\hat{\mathbf{K}}_t - \omega^2_t \hat{\mathbf{M}}_t\right) \boldsymbol{\phi}_t = \boldsymbol{0}$. Here $\hat{\mathbf{K}}_t$ and $\hat{\mathbf{M}}_t$ are the reduced stiffness and mass matrices considering Bloch periodic boundary conditions evaluated at point $t$, and we assume the eigenvector $\boldsymbol{\phi}$ is normalized w.r.t. $\hat{\mathbf{M}}$, i.e., $\boldsymbol{\phi} \cdot \hat{\mathbf{M}}\boldsymbol{\phi} = 1$. Here, opening (or closing) the gap $\Delta \lambda^{(m,n)}_t$ corresponds to opening (or closing) the gap $\Delta \omega^{(m,n)}_t$ in the band diagram, and therefore we equivalently use $\Delta \lambda^{(m,n)}_t$ in the optimization formulations for simplicity.

The second category, i.e., the complete isolation of band $m$ from band $n$ can also be achieved by evaluating the same expression $\Delta \lambda^{(m,n)}_j$ of Eq.~\eqref{eq:gap_function} at all wavevector points $j$ and enforcing them to be non-zero. Ideally, all points along the BZ should be sampled to ensure complete band isolation, but this is computationally prohibitive as each sampled point requires solving an eigenvalue problem. To address this, we enforce a total BG, which is a stricter condition for band isolation, while sampling only at the three HSPs $\Gamma$, M and K for $j$ in the optimization. The BG between bands $m$ and $n$, sampled at the HSPs, can be expressed as
\begin{equation} \label{eq:bandgap_function}
    \Tilde{\Delta} \lambda^{(m,n)} = \dfrac{1}{\tau_\text{KS}[1 / \lambda_j^{(m)}]} - \tau_\text{KS}[\lambda_j^{(n)}], \ \ (m \geq n),
\end{equation}
where $\tau_\text{KS}$ is the KS-aggregation function~\cite{Kreisselmeier1979} given as
\begin{equation}\label{eq:KS}
    \tau_\text{KS}[(\cdot)_j] = \max\limits_{j=\Gamma,\text{K,M}} \{ (\cdot)_j \} + 
        \dfrac{1}{\beta_\text{KS}} \log \left( \sum_{j} e^{\beta(\cdot)_j} \right),
\end{equation}
which provides a smooth approximation of the maximum operator using a smoothness parameter $\beta \in [1,\infty)$. To verify complete spectral isolation, we evaluate the band morphology across all points in the BZ after optimization. %While this computationally convenient strategy, i.e., enforcing BG at HSPs does not explicitly enforce band isolation in the optimization, it can successfully result in band isolation in most cases. Nevertheless, 
%To verify the complete spectral isolation between bands, we post-evaluate their band morphology using all points on the BZ after optimization. % to verify the band isolation after optimization. 

Using Eqs.~\eqref{eq:gap_function} and~\eqref{eq:bandgap_function}, we now construct an optimization problem where the merging of bands 3 and 4 at $\Gamma$ point is promoted through the objective function and all other criteria are realized through different constraint functions. A volume constraint is also included to prevent the possible appearance of redundant materials in the optimized design. The optimization problem is formulated as
\begin{equation} \label{eq:TO_formulation_0}
\begin{array}{ll}
    \min\limits_{\boldsymbol{z}} \ \ & J(\boldsymbol{z}) = \Delta \lambda_{\Gamma}^{(4,3)}(\boldsymbol{z}) \\[5pt]
    \text{s.t.} \ \ & g_1(\boldsymbol{z}) = -\Delta \lambda_{\text{M}}^{(4,3)}(\boldsymbol{z}) + \varepsilon_\text{M} \leq 0, \\[5pt]
                \ \ & g_2(\boldsymbol{z}) = -\Delta \lambda_{\text{K}}^{(4,3)}(\boldsymbol{z}) + \varepsilon_\text{K} \leq 0, \\[5pt]
                \ \ & g_3(\boldsymbol{z}) = -\Tilde{\Delta} \lambda^{(3,2)}(\boldsymbol{z}) + \varepsilon_\text{BG1} \leq 0, \\[5pt]
                \ \ & g_4(\boldsymbol{z}) = -\Tilde{\Delta} \lambda^{(5,4)}(\boldsymbol{z}) + \varepsilon_\text{BG2} \leq 0, \\[5pt]
                \ \ & g_5(\boldsymbol{z}) = V(\boldsymbol{z}) - V^* \leq 0, \\[5pt]
                \ \ & z_e \in [0,1], \quad \quad \quad \quad \quad \quad \quad \quad \ e=1,\ldots,N_e, \\[5pt]
    \text{with} \ \ & \left(\hat{\mathbf{K}}_j(\boldsymbol{z}) - \lambda_j \hat{\mathbf{M}}_j(\boldsymbol{z})\right)\boldsymbol{\phi}_j = \mathbf{0}, \ j=\{\Gamma,\text{M},\text{K}\},
\end{array}   
\end{equation}
where $\boldsymbol{z}$ is the design variable vector representing solid or void regions in the design domain, $V^*$ is the maximum allowable volume fraction, and $\varepsilon_\text{M}, \varepsilon_\text{K}, \varepsilon_\text{BG1}$ and $\varepsilon_\text{BG2}$ are the prescribed constraint tolerances representing lower bounds of desired HSP gaps or BGs. This inverse design process that embeds the insights from TQC into a TO framework for automated discovery of topological lattices is summarized in Fig.~\ref{fig:TopOpt}. Specifically, %where 
Fig.~\ref{fig:TopOpt}A illustrates the periodic design domain that conforms to the crystallographic symmetry and periodicity requirements for the chosen $p31m$ group, Fig.~\ref{fig:TopOpt}B presents the primary design principles for the band morphology and corresponding mathematical expressions in the form of an optimization problem using a simplified representation of Eq.~\eqref{eq:TO_formulation_0}, and Fig.~\ref{fig:TopOpt}C shows the resulting optimized design satisfying all design principles in Fig.~\ref{fig:TopOpt}B, as verified in the next subsection with rigorous numerical and experimental investigations. We note that a unique advantage of using optimization-driven morphogenesis is that, by varying the forms of the objective and constraint functions in Eq.~\eqref{eq:TO_formulation_0}, we can attain a diverse collection of optimized mechanical lattices with dissimilar geometries yet identical topological characteristics that meet the same primary principles. The database of these lattices are presented and discussed in a later section and in the \textit{SI Appendix}. 
%
% Interestingly, interestingly, other variations of the presented optimization formulation~\eqref{eq:TO_formulation_0}, i.e., alteration of objective function and constraints as well as different constraint tolerance values can also be effective in realizing topological characteristics in mechanical lattices as long as the optimized designs satisfy the primary design principles. The database discussed in section~\ref{sec:database} presents different designs with the same topological behavior but optimized with alternate optimization formulations, some of which are included in the SM. 

\begin{figure*}[t!]
\includegraphics[width=0.9\textwidth]{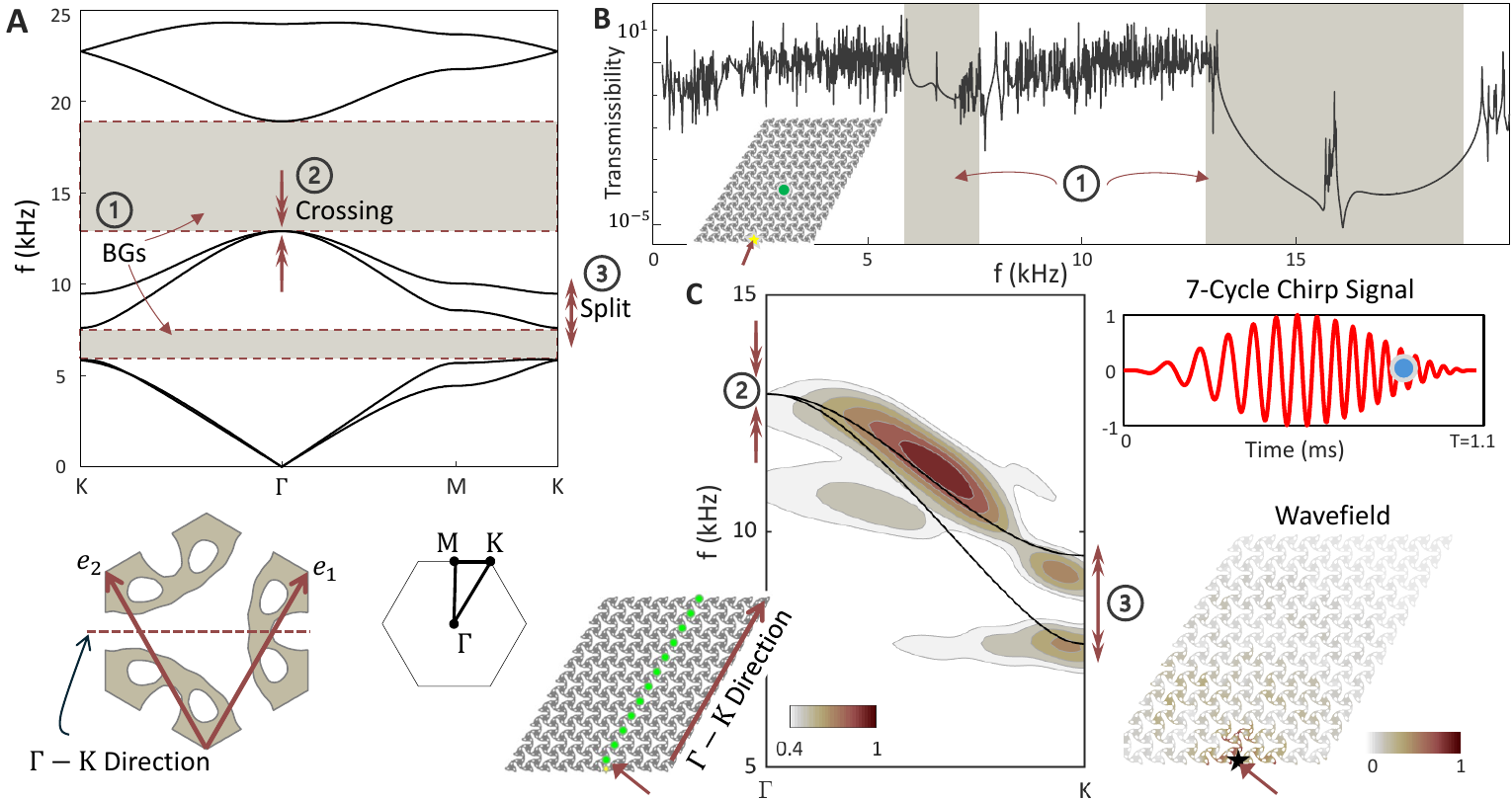}
\caption{Optimized design exhibiting target topological bands. (A) Unit cell derived from the TO algorithm, shown with its calculated band diagram. The mid-frequency bands demonstrate the desired morphological attributes, with the BZ indicated in the inset. (B) Transmissibility versus frequency plot, illustrating the formation of two BGs enclosing finite-frequency passing bands. (C) Band reconstruction from transient simulations using a chirp signal (plotted in red), capturing key morphological features of bands 3-4, specifically the crossing at $\Gamma$ and splitting at K, accompanied by a snapshot of the associated wavefield.}
\label{fig:frag}
\end{figure*}

\subsection{Numerical and Experimental Extraction of Topological Signatures.}
We now proceed to interrogate the outcome of the TO algorithm by performing a full phononic analysis of the configuration that has emerged from the optimization and, specifically, assessing the establishment of topological character. The generated configuration is shown in the inset of Fig.~\ref{fig:frag}A %based on a given set of parameters (LIST PARAMETERS HERE), 
along with its lattice vectors $\mathbf{e}_1$ and $\mathbf{e}_2$.
The unit cell is discretized into a mesh of 4-node iso-parametric 2D plane-stress elements and subjected to a canonical FE-based Bloch analysis(\nameref{method}). The resulting band diagram, limited to the six lowest bands, is shown in Fig.~\ref{fig:frag}A.
%corroborate the findings of the algorithm by analyzing the optimized geometry. 
%The algorithm generated the unit cell (inset of Fig.~\ref{fig:frag}(A)) based on a given set of parameters (LIST PARAMETERS HERE), with the primitive lattice vectors denoted as $\mathbf{e}_1$ and $\mathbf{e}_2$. 
 A visual inspection of the band diagram immediately reveals that the morphological requirements are indeed satisfied: (\romannumeral 1) bands $3-4$ are isolated, (\romannumeral 2) they cross at the HSP $\Gamma$, and (\romannumeral 3) a gap exists at the HSPs M and K (see \textit{SI Appendix} for details on the irreps and mode shapes at the HSPs calculated for this configuration, and the existence of localized modes within the BGs). To corroborate these band characteristics, we conduct full-scale FE simulations on a finite domain obtained via 2D tessellation of the optimized unit cell. Specifically, to capture the first requirement (isolated finite-frequency bands), we perform a frequency-domain simulation under sustained harmonic excitation. The domain is excited at the yellow dot (input) and the response is measured at the green dot (output) of the inset of Fig.~\ref{fig:frag}B. By sweeping the frequency and calculating, for each $\omega$, the output/input ratio, we generate the transmissibility curve shown in Fig.~\ref{fig:frag}B. This curve confirms two attenuation regions (shaded), matching the BG frequencies predicted by Bloch analysis and sandwiching a pass-band region.

To document the second and third features (band crossing pattern), we need to reconstruct the actual morphology of the bands. To this end, we resort to transient simulations to induce waves with a broad wavevector content and we sample them along specific directions corresponding to given intervals of the BZ contour. We excite the domain at the yellow dot (inset of Fig.~\ref{fig:frag}C) with a 7-cycle linear chirp signal polarized perpendicular to the $\Gamma-$K direction, as shown in red in Fig.~\ref{fig:frag}C. The instantaneous frequency is defined as $f(t) = c t+f_0$, where $f_0=6.5$ kHz is the starting frequency, $f_{1}=26$ kHz is the final frequency, $c=\frac {f_{1}-f_{0}}{T}$ is the chirp rate, and $T=1.1$ ms is the total sweep time. A snapshot of the wavefield at the time corresponding to the blue point in the signal plot is also shown in Fig.~\ref{fig:frag}C, with colormap corresponding to the normalized displacement magnitude. We collect displacement time histories at discrete, %12
equally spaced points along $\Gamma-$K (green dots in Fig.~\ref{fig:frag}C inset), and perform 2D DFT to transform the spatio-temporal dataset into a frequency-wavenumber dataset. The resulting spectral amplitude contours, overlaid on the band structure in Fig.~\ref{fig:frag}C, confirm the targeted morphological characteristics, with a single prominent spectral feature at $\Gamma$ and, in contrast, a split signature at K with two spectral features conforming to the limits of bands 3 and 4, respectively. 

%Notably, %in the $p31m$ symmetry class, 
%the HSP M only admits 1D representations ($M_{1(2)}$), ensuring that the gap at M is automatically satisfied. 

\begin{figure*}[t!]
\includegraphics[width=0.9\textwidth]{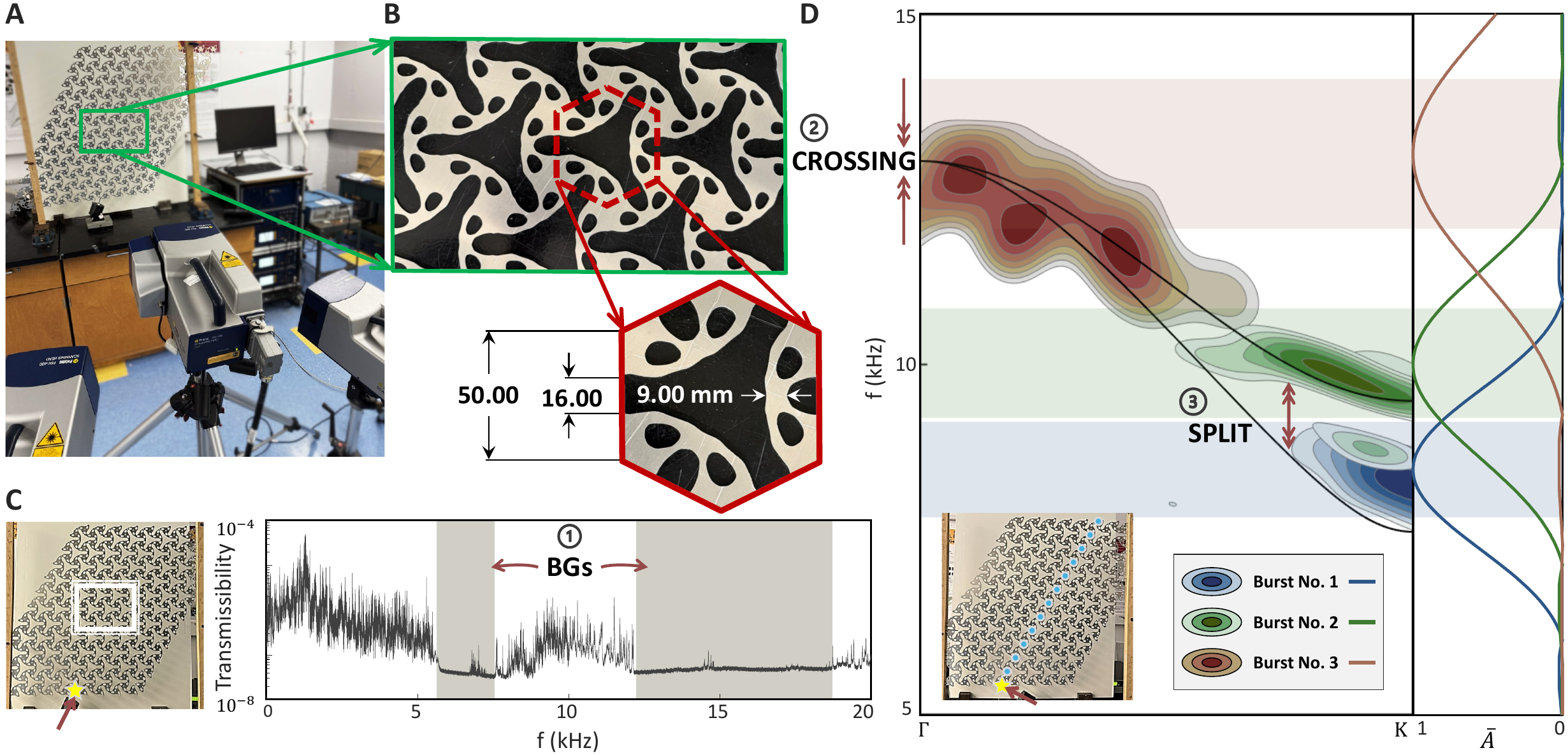}
\caption{\label{fig:exp}Experiments confirming the emergence of topological signatures: prototype of a fabricated lattice following result from TO algorithm. (A and B) Experimental setup for 3D SLDV testing and a close-up view of the specimen, with the geometric details of its corresponding unit cell shown in the inset. (C) Experimental transmissibility curve revealing finite-frequency pass-band between BGs. (D) Experimentally reconstructed band morphology featuring crossing at $\Gamma$ and split at K, revealing signatures of nontrivial topology according to TQC irreps. The spectra of the excitation signals are shown on the right, with color-coded regions indicating the frequency intervals of highest activation for each tone.}
\end{figure*}

%\textbf{Experimental Evidence of Topological States.} 
To confirm the robustness of the established topological signatures of the optimized configuration in transitioning from ideal simulations to physical implementations, we now complement our numerical predictions with experimental validation using laser vibrometry (Polytec PSV 400 3D Scanning Laser Doppler Vibrometer (SLDV)) on a physical prototype. The experimental setup is illustrated in Fig.~\ref{fig:exp}A, with a zoomed-in view of the waterjet-cut prototype geometry shown in the inset of Fig.~\ref{fig:exp}B (all details provided in \nameref{method}). The experiments confirm the emergence of the key dynamical features associated with topological bands at finite frequencies for this symmetry class. First, we apply a broadband pseudo-random excitation at the point marked by the yellow dot in Fig.~\ref{fig:exp}C, and we measure the in-plane velocity at selected sampling points within the white box. The experimentally reconstructed transmissibility curve reveals the opening of two BGs, confining a pass-band region, with excellent qualitative and satisfactory quantitative match of its numerical counterpart. Next, in Fig.~\ref{fig:exp}D, we experimentally reconstruct the bands 3-4 morphology by exciting the frequency range of the bands using three concatenated 7-cycle narrow-band tone burst excitations with carrier frequencies of 8.5 kHz, 10 kHz, and 13 kHz, and sampling the response along the relevant lattice direction. The spectral amplitude maps conform to the bands predicted by Bloch analysis, confirming the crossing at $\Gamma$ and the split at K. The FFT spectra of the tone bursts, shown in the right panel of Fig.~\ref{fig:exp}D, are color-coded to highlight the corresponding frequency intervals with the top $40\%$ activation.
%First, we aim to experimentally demonstrate the presence of two BGs. To achieve this, we apply a broadband pseudo-random excitation at the point marked by the yellow dot in Fig.~\ref{fig:exp}(C). The in-plane velocity is measured at designated sampling points within the white box (Fig.~\ref{fig:exp}(C)), and the resulting transmissibility versus frequency curve is plotted in the same figure. Two distinct regions of attenuation are observed, confirming the existence of finite-frequency isolated bands in the mid-frequency spectrum. Second, we investigate the morphological features of the intermediate bands by analyzing the spectral response to narrowband excitations. We collect time histories of the in-plane velocity at evenly spaced points along the $\Gamma$-$K$ direction (blue dots in Fig.~\ref{fig:exp}(D) inset) under 7-cycle narrowband tone-burst excitations with carrier frequencies of 8.5 kHz, 10 kHz, and 13 kHz. The resulting 2D DFT spectral amplitude contours are shown in Fig.~\ref{fig:exp}(D), with each burst represented in a different color. These contours confirm the targeted morphological attributes, as the blobs align with the expected band diagram. The FFT spectra of the tone bursts are plotted on the right panel of Fig.~\ref{fig:exp}(D), color-coded with the highlighted regions which correspond the frequency intervals with the $\%40$ highest activation.

\subsection{Design Space Exploration and Construction of Configuration Libraries.}\label{sec:database}
\begin{figure*}[t!]
\centering
\includegraphics[width=0.9\textwidth]{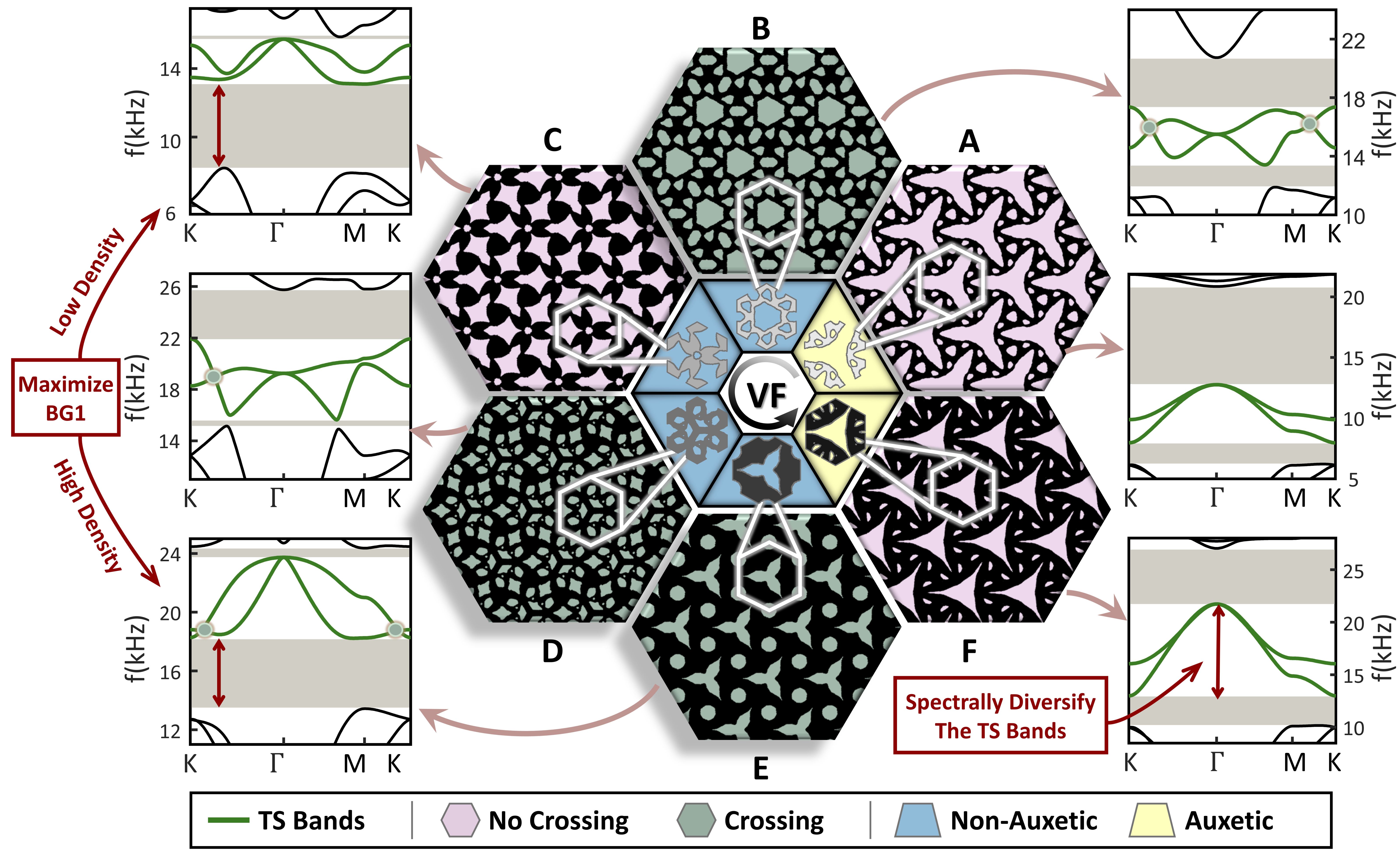}
\caption{\label{fig:library}A diverse library of discovered metamaterials that share common topological attributes, illustrated using results for the benchmark symmetry class. The optimized unit cells and their tessellated configurations are organized by VF. The corresponding band diagram for each case is presented within the frequency range of the TSs (green lines). All cases are color-coded to indicate the presence or absence of additional crossings at wave vectors along the high-symmetry lines. Cases A and F are auxetic, while the others have a positive Poisson’s ratio and are shaded accordingly. Although the designs retain the same topological character due to enforced band morphology constraints, they exhibit distinct geometric variations.}
\end{figure*}
Next, we explore the design space to identify alternative configurations that belong to the same symmetry class and preserve the band morphology characteristics necessary for the existence of topologically nontrivial states. The broader objective of this design exploration is to demonstrate the capability of this framework to discover, as a result of a simple parameter calibration, an entire library of configurations that share common topological attributes and simultaneously achieve a plethora of secondary phononic characteristics. The chart in Fig.~\ref{fig:library} illustrates a collection of optimized configurations obtained through parameter alteration introduced in the TO formulation that result in domains with different volume fractions (VFs) (either by design, by imposing desired bounds on VF as constraints, or as a spontaneous byproduct of the algorithm targeting some functionalities objectives). %The specifics of the algorithm variants are given in the SM. 
Six optimized unit cells are shown in six trapezoidal boxes organized counterclockwise according to their VF. For each configuration, the resulting tessellation is shown in the corresponding hexagonal sector of the chart to highlight the variety of patterns and connectivity landscapes that can be achieved. 
For each case, we report a portion of the corresponding band diagram corresponding to the frequency range that encompasses the TSs (highlighted) to emphasize that in all cases the common morphological band features enforced by the algorithm are indeed met.
The explored variations include adjusting design parameters such as BG width and VF, swapping objectives and constraints, or imposing additional constraints. See \textit{SI Appendix} on the specific TO formulations used in each case. 
%For instance, in Case A, the objective is to maximize the BG while shifting the original objective function into the constraints, along with other existing constraints. Additionally, an upper bound constraint is imposed on the VF. Case B, in contrast, follows a TO formulation similar to the original optimized unit cell presented in the previous section, but allows for a higher upper bound on VF. 

Interestingly, even slight modifications to the TO formulation lead to significant changes in both the real-space geometry and the reciprocal-space band structure due to the highly non-convex nature of the mathematical optimization problem, with influences on the phononic or mechanical properties. An intriguing occurrence is the appearance of additional band crossings along the high-symmetry lines $\Gamma-$K and/or M$-$K. This feature is observed in Cases B, D, and E. Cases featuring crossing are highlighted in Fig.~\ref{fig:library} by green backgrounds in the tessellations against pink backgrounds for the regular cases. Notably, these crossings not occurring at the HSPs do not alter the topological character of the bands. Therefore, these cases are correctly obtained by the algorithm. It is also interesting to see how different configurations can be classified according to their bulk, shear moduli, and Poisson's ratio. These quantities can be inferred from a manipulation of the wave speeds of the acoustic modes in the long-wavelength limit, as discussed in~\cite{phani2006wave}. Remarkably, Cases A and F, which have the most kagome-like geometries, exhibit auxetic behavior (shaded light yellow in Fig.~\ref{fig:library}), whereas the other cases have a positive Poisson’s ratio (shaded blue in Fig.~\ref{fig:library}). See the \textit{SI Appendix} for more details. Additional differences between these outcomes can be pointed out that can be traced to differences in the TO algorithm. Cases C and E share similar TO formulations, both aiming to maximize BG1 while maintaining a similar lower bound constraint on VF. However, in Case C, the algorithm allows lower values for BG2, ultimately yielding two design options: one with a lower effective mass density (C) and another with a higher effective mass density (E). Case D is designed to simultaneously maximize the gaps at the HSPs K and M between bands 2–3 and 4–5, while enforcing all other requirements as constraints. Lastly, Case F is designed to enhance experimental feasibility by optimizing the TSs for maximum bandwidth. In other words, we increase the spectral separation between the peak and valley frequencies of the TSs to facilitate experimental validation and simplify testing.

\section{Concluding Remarks}
TO has emerged as a powerful tool for designing metamaterials with tailored mechanical and wave manipulation properties. While significant progress has been made in optimizing conventional performance metrics—such as stiffness, toughness, and BG width—the systematic integration of topological constraints into TO frameworks remains largely unexplored. This gap stems from the inherent challenge of mathematically formulating topological requirements in a rigorous and computationally tractable manner. In this work, we have addressed these challenges by introducing a methodology that seamlessly integrates band topology criteria into TO algorithms. Leveraging the framework of TQC, we have identified key band morphological attributes that serve as reliable indicators of topological character. Beyond its theoretical contributions, this study enables the automated generation of metamaterials with on-demand topological properties, bridging the gap between theoretical predictions and experimental realization. Additionally, our findings provide a comprehensive library of lattice configurations, each engineered to satisfy a complementary wave manipulation property while constrained to the same primary topology, thereby establishing a structured approach to metamaterial discovery. Looking ahead, the implications of this work can be extended beyond the studied symmetry group, as the developed methodologies can be adapted to other symmetry classes, where further considerations need to be incorporated, yet the band morphology plays a similarly fundamental role. %This research marks a significant step toward the integration of topological mechanical metamaterials into real-world applications, ultimately expanding the design space for next-generation wave control technologies.

%Topology optimization has been successfully applied to the design of structural materials and metamaterials, particularly for stiffness and bandgap optimization. However, the integration of topological mechanical metamaterial properties within TO frameworks remains largely unexplored, with previous efforts being problem-specific and lacking a unified methodology. This study advances the field by systematically incorporating topological attributes into TO algorithms, leveraging the TQC framework to guide the optimization process. The availability of symmetry-based topological databases, such as the Band Connectivity Server BCS, enables a more rigorous and automated approach to designing topological phononic materials. Our configuration library further demonstrates that, while maintaining the primary constraint of belonging to the same topological class, a wide range of configurations can be realized to accommodate various complementary design objectives. %This work establishes a foundation for the rational design of topological metamaterials, expanding their applicability in wave control and elastic systems. 

\section{\label{method} Materials and Methods}

\subsection{Simulation and Optimization Setup}
We generate a quadrilateral isoparametric mesh with approximately 14,000 elements within a hexagonal unit cell domain, where each side is set to unity (Fig.~\ref{fig:TopOpt}A). The mesh is refined appropriately while preserving the targeted symmetry group, $C_{3v}$. To construct the symmetric mesh, we first discretize the highlighted region in Fig.~\ref{fig:TopOpt}A using GMSH. We then apply the mirror symmetry operation $m_{11}$, reflecting the mesh across the plane perpendicular to $\mathbf{e}_1 +\mathbf{e}_2$. Subsequently, we apply two successive rotations, $C_3^1$ and $C_3^2$, to generate a symmetric mesh across the entire domain. Details of the TO, including design parameterization with filtering~\cite{Bruns2001, Bourdin2001, Sigmund2007,Talischi2012} and Heaviside projection~\cite{Wang2010}, interpolation rules, and optimization formulations alternative to Eq.~\eqref{eq:TO_formulation_0} are provided in \textit{SI Appendix}.

\subsection{Material Properties and Optimized Unit Cell Specification}
The optimization process yields a pixelated black-and-white hexagon unit cell (Fig.~\ref{fig:TopOpt}C), where black and white regions correspond to solid and void, respectively. Due to the jagged nature of the boundaries, we import the geometry into SolidWorks and smooth the edges using splines. The smoothed domain is then scaled so that each side of the hexagon measures 50 mm. To construct a FE- model, the refined geometry is exported as a .STEP file and imported into GMSH, where it is discretized into approximately 2,880 elements for band diagram computation using Bloch periodic boundary conditions. For full-scale simulation analysis, we tessellate the unit cell to construct a finite parallelogram domain consisting of 120 unit cells. For experimental testing, the finite domain geometry is exported as an .STL file for manufacturing. Figs.~\ref{fig:exp}A and B show the fabricated specimen, produced via water-jet cutting from a 2-mm-thick aluminum sheet, which is vertically constrained via boundary supports. The material properties of aluminum are: Young's modulus = 71 GPa, Poisson’s ratio = 0.33, and mass density = 2,700 $kg/m^3$. 

\subsection{The 3D Laser Doppler Vibrometer Experiments.}

The Polytec PSV 400 3D Scanning Laser Doppler Vibrometer (SLDV), equipped with three scanning laser heads, is used to acquire in-plane velocity measurements at predefined scan points. Excitation is applied via an electromechanical shaker (Brüel \& Kj\ae r Type 4810), which is internally triggered by the vibrometry setup through an amplifier (Brüel \& Kj\ae r Type 2718). The shaker excites the lattice through a stinger at the desired location. Retro-reflective tape is applied at the scan points to enhance reflectivity and reduce noise in the data. The acquired velocity data are decomposed into $\hat{x}$, $\hat{y}$, and $\hat{z}$ components using Euler angles, which are internally computed by the PSV software as part of the 3D alignment process. This data is further processed in MATLAB to construct the transmissibility curve and reconstruct wavefields and DFT plot.

To experimentally construct the transmissibility curve, we apply the excitation signal in the 0–20 kHz range in three subranges: 0–5 kHz, 5–10 kHz, and 10–20 kHz. To ensure sufficient energy is supplied at higher frequencies, we prescribe progressively higher amplitudes for each successive range, allowing for a more uniform energy injection across the entire spectrum. To reconstruct the TSs along the $\Gamma-$K direction, we note that this direction aligns with the lattice vector $\mathbf{e}_1$ (inset of Fig.~\ref{fig:frag}A). Therefore, we measure the time histories of the $\hat{x}$ and $\hat{y}$ in-plane velocity components at 12 equally spaced scan points along the $\Gamma-$K direction and separated by the magnitude of the primitive lattice vector $\mathbf{e}_1$ (blue dots in the inset of Fig.~\ref{fig:exp}D). To mitigate the influence of low-frequency ambient vibrations, we apply a high-pass filter embedded in the vibrometry software. The resulting spatiotemporal data are then subjected to a 2D-DFT to obtain the spectral amplitude contours shown in Fig.~\ref{fig:exp}D.

\subsection{Acknowledgment}
P.A. acknowledges the support of the UMN CSE Graduate Fellowship. P.A. and S.G. acknowledge support from the National Science Foundation (grant CMMI-2027000 and CMMI-2344257). X.S.Z., R.D.K., and W.L. acknowledge support from the Air Force Office of Scientific Research (award FA9550-23-1-0297) and National Science Foundation (grant CMMI-2344258) and K.S. acknowledges the support from the Office of Naval Research (grant MURI N00014-20-1-2479).

%\bibliographystyle{apsrev4-1}
%\bibliography{ref}% Produces the bibliography via BibTeX.

%\input{refbbl.bbl}
%\bibliographystyle{naturemag}
%\bibliography{ref}% Produces the bibliography via BibTeX.
\bibliographystyle{apsrev4-1}
\bibliography{References}

\onecolumngrid

%%%%%%%%%%%%%%%%%%%%%%%%%%%%%%%%%%%%%%
%%   Supplementary Information
%%%%%%%%%%%%%%%%%%%%%%%%%%%%%%%%%%%%%%
\section*{\Large\bf Supplemental Material}
%\counterwithout{figure}{section} 
\makeatletter

\renewcommand \thesection{S-\@arabic\c@section}
\renewcommand\thetable{S\@arabic\c@table}
\renewcommand{\thefigure}{S\arabic{figure}}
\renewcommand \theequation{S\@arabic\c@equation}
\makeatother
\setcounter{equation}{0}  %  this will re-count eq from 1
\setcounter{figure}{0}  %  this will re-count eq from 1
\setcounter{section}{0}  %  this will re-count eq from 1
%\counterwithin{figure}{section}

\maketitle
\section{Topological State (TS)-informed Topology Optimization (TO)}
This section presents the relevant technical details about the TS-informed TO framework for the automated discovery of topological mechanical metamaterials. In the next subsections, we discuss the design parameterization, interpolation rules, and some optimization formulations that are used to generate the design database presented in the main article. 

\subsection{Design Parameterization and Interpolation Schemes}
We adopt the design parameterization of the standard density-based TO~\cite{Bendsoe2004_TObook,Wang2021}. In this method, a density field $z(\boldsymbol{x})$ characterizes the structural geometry, i.e., the solid-void distribution in our 2D hexagonal design domain $\Omega(\boldsymbol{x})$, with $\boldsymbol{x}$ denoting the position of any point in the domain. For computational purposes, we adopt a piece-wise distribution of $z(\boldsymbol{x})$ as $\boldsymbol{z}=\{z_1,\ldots,z_{N_e}\}$, where $z_e$ denotes the density variable for element $e$ in the design domain $\Omega(\boldsymbol{x})$ discretized with $N_e$ number of Q4 finite elements. We apply a density filter~\cite{Bruns2001, Bourdin2001, Sigmund2007} and a $p31m$ wallpaper group symmetry mapping on the design variable $\boldsymbol{z}$ sequentially to regularize the design space and enforce the symmetry conditions of $p31m$ wallpaper group lattice class, respectively. The filtered variable $\tilde{\boldsymbol{z}}$ is obtained as $\tilde{z}_{k} = P^{(s)}_{ki} \ P^{(d)}_{ij} z_{j}$, where the symmetry mapping $\mathbf{P}^{(s)}$ is constructed according to the mirror and rotation symmetry rules of $p31m$ wallpaper group lattice class as elaborated in the \textit{Materials and Methods} section of the main article, and the density filter mapping $\mathbf{P}^{(d)}$ is expressed as~\cite{Talischi2012} 
\begin{equation} \label{eq:filter}
\begin{array}{ll}
     \ P^{(d)}_{ij} = \dfrac{ w_{ij} v_{j}} {\sum\limits_{k=1}^{N_e} w_{ik} v_{k}}, \quad \text{with}, \ w_{ij}=\text{max}\left\{0,1-\dfrac{\left\|\boldsymbol{x}_{i}-\boldsymbol{x}_{j}\right\|}{R_z}\right\}, 
\end{array}
\end{equation}
where $\left\|\boldsymbol{x}_{i}-\boldsymbol{x}_{j}\right\|$ is the distance between the centroids ${x}_{i}$ and ${x}_{j}$ corresponding to elements $i$ and $j$, respectively, $v_{j}$ is the volume of element $j$, and $R_z$ is the filter radius. To obtain a discrete design at the end of the optimization, we use a Heaviside projection~\cite{Wang2010} given by
\begin{equation} \label{eq:Heaviside}
    \overline{z}_{e}=\frac{\tanh (\beta \eta)+\tanh (\beta(\Tilde{z_{e}}-\eta))}{\tanh (\beta \eta)+\tanh (\beta(1-\eta))},
\end{equation}
where, $\beta$ is the Heaviside projection sharpness parameter, $\eta$ is the Heaviside projection threshold, and $\boldsymbol{\overline{z}}$ is the physical density variable that characterizes the design with $\overline{z}_e = 1$ and $\overline{z}_e = 0$ representing solid and void, respectively, for element $e$. In this study, we use $\eta=0.5$ throughout the optimization, and gradually increase $\beta$ from 1 to 128 by doubling it every 50 iterations.

 For an element with $\overline{z}\in[0,1]$, we use linear interpolations for stiffness (equivalently for Young's modulus, assuming similar Poisson's ratios for solid and void elements) $E$ and mass density $\rho$ as
\begin{equation} \label{eq:interpolation_rules}
\begin{array}{ll}
    E(\overline{z}_e)=\left[\varepsilon_E + (1-\varepsilon_E)\overline{z}\right]E^0 \\ [5pt]
    \rho(\overline{z}_e)=\left[\varepsilon_\rho + (1-\varepsilon_\rho)\overline{z}\right]\rho^0 \\
\end{array}
\end{equation}
where $\varepsilon_E$ and $\varepsilon_\rho$ are small numbers to avoid numerical singularity, and they represent the relative stiffness and mass density of void elements. In this study, we use $\varepsilon_E=10^{-5}$ and $\varepsilon_\rho=10^{-6}$.% to eliminate void vibration modes from the lowest six eigenfrequencies used for band analysis. %For the solid material properties, we assume $E^0=71~\text{GPa}$, $\nu^0=0.33$, and $\rho^0=2,700~\text{kg/m}^3$ for Young's modulus, Poisson's ratio, and mass density, respectively. 

% as in the SIMP method~\cite{Zhou1991,Bendse1999}

%\subsection{Bloch periodic boundary conditions for band analysis}
% include analysis details -- i.e., construction of W matrix, and transformation of stiffness & mass matrices

\subsection{Alternative Optimization Formulations}
The main article presents an optimization formulation (Fig.~2) that successfully generates an optimized design with desired topological properties. Here, we include some alternative optimization formulations that are used to generate the design database in Fig.~5. The alterations are driven by different additional desired features in a topological design, as listed in Table~\ref{tab:formulations}. These formulations do not strictly restrict the volume usage in the design, and therefore allow the emergence of a wide range of additional features.

{\renewcommand{\arraystretch}{1.4}
\begin{table}[htp]
\caption{Alternative optimization formulations for various desired features in topological metamaterial}
    \centering
    \begin{tabular}{p{4cm}<{\centering} p{4cm}<{\centering} p{4cm}<{\centering} p{5cm}<{\centering}}
    \hline
    \hline
       \begin{tabular}[c]{@{}c@{}}Desired \\ additional \\features\end{tabular}  & \begin{tabular}[c]{@{}c@{}}Enlarge bandgap (BG)\\ between bands 2-3\end{tabular} & \begin{tabular}[c]{@{}c@{}}Spectrally diversify\\ bands 3 and 4 \end{tabular} & \begin{tabular}[c]{@{}c@{}}Promote equal \\ BGs between \\ bands 2-3 and band 4-5\end{tabular}\\
       \hline
       \hline
       \begin{tabular}[c]{@{}c@{}} minimize: \end{tabular} 
       & $-\Tilde{\Delta} \lambda^{(3,2)}(\boldsymbol{z})+\alpha_v V(\boldsymbol{z})$
       & $V(\boldsymbol{z})$ 
       & \begin{tabular}[c]{@{}c@{}}$\max\{-\Tilde{\Delta} \lambda^{(3,2)}(\boldsymbol{z}), -\Tilde{\Delta} \lambda^{(5,4)}(\boldsymbol{z})\}$ \\+$\alpha_v V(\boldsymbol{z})$\end{tabular}\\
       \hline
       \begin{tabular}[c]{@{}c@{}} subject to: \end{tabular}
           & \(
            \begin{array}{ll}
                \Delta \lambda_{\text{M}}^{(4,3)}(\boldsymbol{z}) \geq \varepsilon_\text{M}, \\[5pt]
                \Delta \lambda_{\text{K}}^{(4,3)}(\boldsymbol{z}) \geq \varepsilon_\text{K}, \\[5pt]
                \Delta \lambda_{\Gamma}^{(4,3)}(\boldsymbol{z}) \geq \varepsilon_{\Gamma}, \\[5pt]
                \Tilde{\Delta} \lambda^{(5,4)}(\boldsymbol{z}) \geq \varepsilon_\text{BG2}, \\[5pt]
                V(\boldsymbol{z}) \geq V^*, \\
            \end{array}   
        \) & \(
            \begin{array}{ll}
                \Delta \lambda_{\text{M}}^{(4,3)}(\boldsymbol{z}) \geq \varepsilon_\text{M}, \\[5pt]
                \Delta \lambda_{\text{K}}^{(4,3)}(\boldsymbol{z}) \geq \varepsilon_\text{K}, \\[5pt]
                \Delta \lambda_{\Gamma}^{(4,3)}(\boldsymbol{z}) \geq \varepsilon_{\Gamma}, \\[5pt]
                \Tilde{\Delta} \lambda^{(3,2)}(\boldsymbol{z}) \geq \varepsilon_\text{BG1}, \\[5pt]
                \Tilde{\Delta} \lambda^{(5,4)}(\boldsymbol{z}) \geq \varepsilon_\text{BG2}, \\[5pt]
                \Tilde{\Delta} \lambda^{(3,3)}(\boldsymbol{z}) \geq \varepsilon_\text{BW3}, \\[5pt]
                \Tilde{\Delta} \lambda^{(4,4)}(\boldsymbol{z}) \geq \varepsilon_\text{BW4}, \\[5pt]
                V(\boldsymbol{z}) \geq V^*, \\
            \end{array}   
        \) & \(
            \begin{array}{ll}
                \Delta \lambda_{\text{M}}^{(4,3)}(\boldsymbol{z}) \geq \varepsilon_\text{M}, \\[5pt]
                \Delta \lambda_{\text{K}}^{(4,3)}(\boldsymbol{z}) \geq \varepsilon_\text{K}, \\[5pt]
                \Delta \lambda_{\Gamma}^{(4,3)}(\boldsymbol{z}) \geq \varepsilon_{\Gamma}, \\[5pt]
                \Tilde{\Delta} \lambda^{(3,2)}(\boldsymbol{z}) \geq \varepsilon_\text{BG1}, \\[5pt]
                \Tilde{\Delta} \lambda^{(5,4)}(\boldsymbol{z}) \geq \varepsilon_\text{BG2}, \\[5pt]
                V(\boldsymbol{z}) \geq V^*, \\
            \end{array}   
        \)
        \\
        \hline
       \begin{tabular}[c]{@{}c@{}}Example\\ parameter \\values\end{tabular} 
       & \(
            \begin{array}{cc}
                \alpha_v=0.001, \\[5pt]
                \varepsilon_\text{M}=0.05, \\[5pt]
                \varepsilon_\text{K}=0.05, \\[5pt]
                \varepsilon_{\Gamma}=0.01, \\[5pt]
                \varepsilon_\text{BG2}=0.1, \\[5pt]
                V^*=0.25, \\
            \end{array}   
        \) & \(
            \begin{array}{cc}
                \varepsilon_\text{M}=0.2, \\[5pt]
                \varepsilon_\text{K}=0.2, \\[5pt]
                \varepsilon_{\Gamma}=0.01, \\[5pt]
                \varepsilon_\text{BG1}=0.3, \\[5pt]
                \varepsilon_\text{BG2}=1.0, \\[5pt]
                \varepsilon_\text{BW3}=0.5, \\[5pt]
                \varepsilon_\text{BW4}=0.3, \\[5pt]
                V^*=0.3, \\
            \end{array}   
        \) & \(
            \begin{array}{cc}
                \alpha_v=0.001, \\[5pt]
                \varepsilon_\text{M}=0.075, \\[5pt]
                \varepsilon_\text{K}=0.075, \\[5pt]
                \varepsilon_{\Gamma}=0.01, \\[5pt]
                \varepsilon_\text{BG1}=0.1, \\[5pt]
                \varepsilon_\text{BG2}=0.1, \\[5pt]
                V^*=0.25, \\
            \end{array}   
        \) \\
    \hline
    \hline
    \multirow{1}{*}{BW: Bandwidth, i.e., the spectral separation between the peak and valley frequencies of a single band}
    \end{tabular}
    \label{tab:formulations}
\end{table}
}
%\ \ & z_e \in [0,1], \\[5pt]
%\text{with} \ \ & \left(\hat{\mathbf{K}}_j(\boldsymbol{z}) - \lambda_j \hat{\mathbf{M}}_j(\boldsymbol{z})\right)\boldsymbol{\phi}_j = \mathbf{0},

%\subsection{Feasibility of the restricted/regularized solution space}
% desired property should be realizable (existence of a solution)
% to extreme properties, initial guess with relaxed criteria may help
% design behavior and achieving solutions to extreme criteria may also depend on design resolution; finer mesh with a smaller filter radius may help in this aspect, at the expense of larger computational cost.

\section{Irreducible Representations at the High-Symmetry Points of the Brillouin Zone}

\begin{figure}[t!]
\includegraphics[width=0.9\columnwidth]{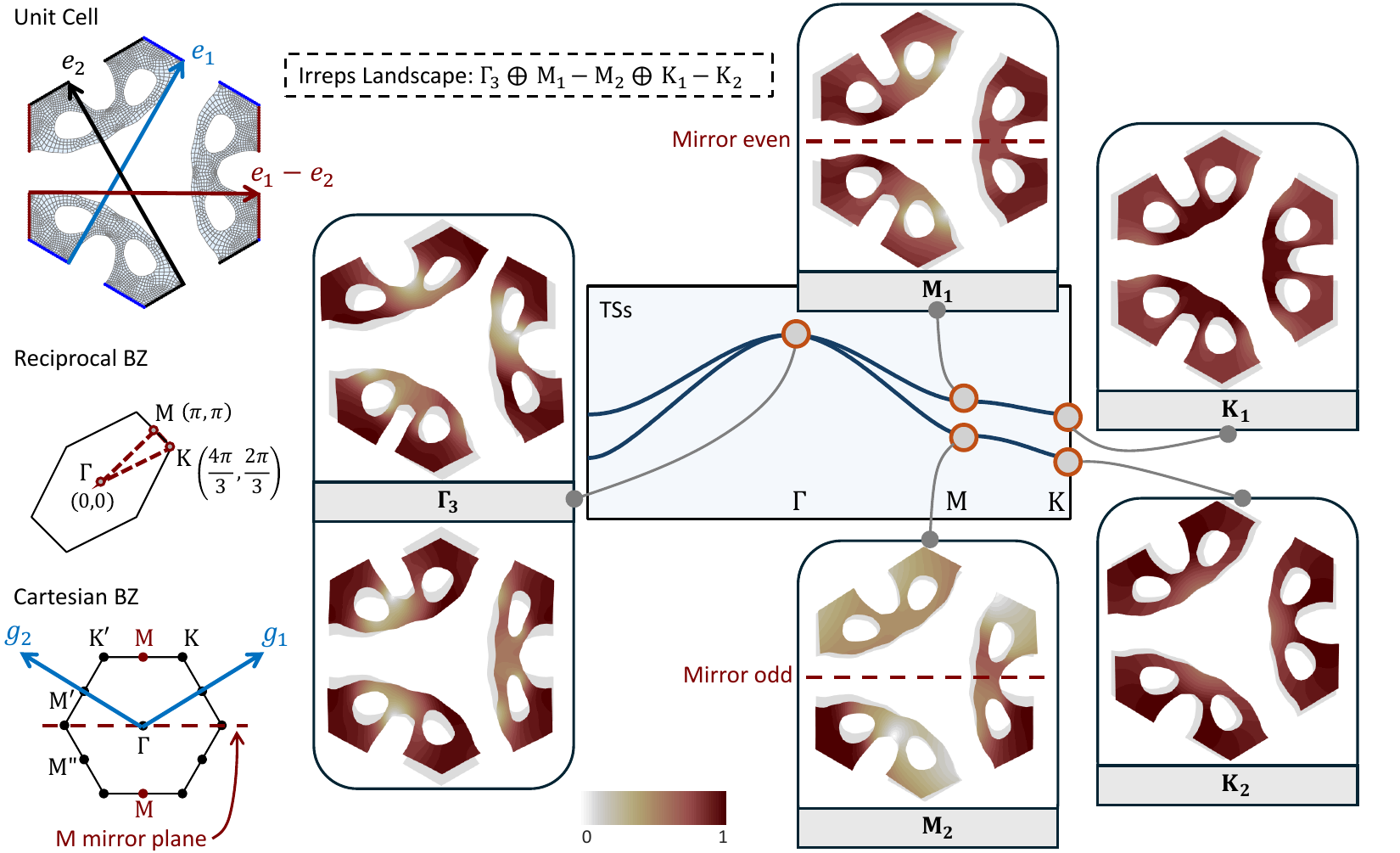}
\caption{\label{fig:irreps}Irreps at the HSPs of the BZ. The discretized unit cell with Bloch nodes color-coded with respect to the lattice vectors is shown. The BZ is illustrated in both reciprocal and Cartesian coordinates. Mode shapes and their corresponding irreps at the HSPs of the TSs are plotted. The color bar represents displacement magnitude, normalized by the highest displacement in each panel.}
\end{figure}

In this section, we compute the irreducible representations (irreps) of the third and fourth bands at the high-symmetry points (HSPs) of the Brillouin zone (BZ) to confirm their topological nature.  Fig.~\ref{fig:irreps} shows the FE- model of the unit cell, which belongs to the $p31m$ space group. The lattice and reciprocal lattice vectors are defined as:
\begin{equation}
\begin{split}
\mathbf{e}_1 &= a\left(\frac{\sqrt{3}}{2}\mathbf{\hat{x}}+\frac{3}{2}\mathbf{\hat{y}}\right)\\
\mathbf{e}_2 &=a\left(-\frac{\sqrt{3}}{2}\mathbf{\hat{x}}+\frac{3}{2}\mathbf{\hat{y}}\right)\\
\mathbf{g}_1 &= \frac{2\pi}{3a}\left(\sqrt{3}\mathbf{\hat{x}}+\mathbf{\hat{y}}\right)\\
\mathbf{g}_2 &= \frac{2\pi}{3a}\left(-\sqrt{3}\mathbf{\hat{x}}+\mathbf{\hat{y}}\right)\\
\end{split}
\end{equation}
such that $\mathbf{e}_i \cdot \mathbf{g}_j = 2\pi \delta_{ij}$. Expressed in the basis of $\mathbf{g}_1$ and $\mathbf{g}_2$, the HSPs are given by $\Gamma = (0,0)$, K $= (2/3,1/3)$, and M $= (1/2,1/2)$. In this symmetry group, the little co-groups at $\Gamma$, K, and M are $C_{3v}$, $C_{3v}$, and $C_{s}$, respectively. The eigenfunctions at these HSPs transform according to the irreps of the corresponding little co-group. From Fig.~\ref{fig:irreps}, we observe that the modes at $\Gamma$ are degenerate, forming the $\Gamma_3$ representation. At M, the modes transform under M$_1$ or M$_2$ depending on their parity with respect to the $m_{11}$ mirror (dashed line), which leaves M invariant as denoted in the Cartesian BZ. The third (fourth) band's mode shape at K is invariant under $C_3$ but odd (even) under the horizontal mirror, forming the K$_2$ (K$_1$) representation. Thus, the degenerate eigenfunctions at $\Gamma$ transform under the 2D irrep $\Gamma_3$, while those at M and K transform under 1D irreps M$_2$, M$_1$ and K$_2$, K$_1$, respectively. According to the Bilbao Crystallography Server (BCS)~\cite{aroyo_et_all_Billcrys_BulgChemCommun_2011,Aroyo_et_all_Billcrys_2006I,Aroyo_et_all_Billcrys_2006II} this irreps landscape $(\Gamma_3\oplus \text{M}_1-\text{M}_2\oplus \text{K}_1-\text{K}_2)$ cannot be described by any symmetric, exponentially localized Wannier functions (SLWF), indicating the topological nature of these bands. Notably, among all elementary band representations constructed from SLWF and tabulated in BCS, no isolated bands exist for wallpaper group $p31m$ (with time reversal symmetry) that exhibit degeneracy at $\Gamma$ but not at K. Furthermore, this irreps configuration is the only way to allow degeneracy at $\Gamma$ and no degeneracy at K for a pair of bands, as permitted by compatibility relations. Therefore, this band morphology necessarily corresponds to this specific irreps landscape and thus guarantees topological bands.

\section{Existence of Localized Modes Within the BGs Bounding the TSs}
\begin{figure}[t!]
\includegraphics[width=0.5\columnwidth]{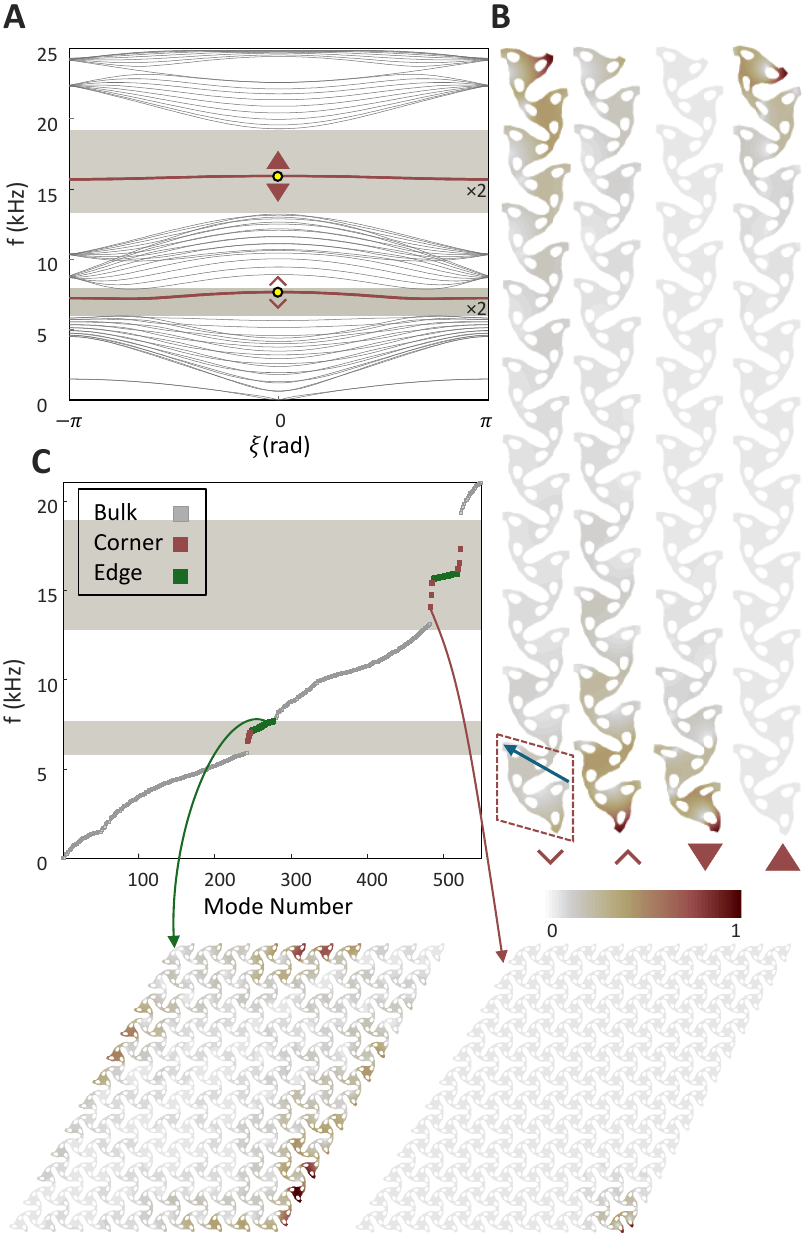}
\caption{\label{fig:local}Localized states within the optimized supercell and the finite domain configuration. (A and B) Band diagram of a 10-cell optimized supercell, showing selected mode shapes of degenerate edge state pairs sampled at $\xi=0$, indicated by corresponding markers within each BG. Supercell unit cell is highlighted with dashed lines, with the Bloch periodic boundary condition applied along the direction of the blue arrow. (C) Eigenfrequency plot for a finite domain configuration, color-coded to distinguish bulk, edge, and corner modes, with two representative eigenfields highlighting an edge mode within the first BG and a corner mode within the second.}
\end{figure}

In this section, we investigate the presence of localized modes within the BG that bound the TSs. To this end, we perform a supercell analysis on a 10-cell supercell, with the unit cell highlighted in Fig.~\ref{fig:local}B. The Bloch periodic boundary condition is applied along the direction indicated by the blue arrow, while the opposite boundary remains free. The resulting band diagram, shown in Fig.~\ref{fig:local}A, reveals a pair of relatively flat, degenerate bands within each BG (shaded regions), represented by brown lines. Additionally, we extract the mode shapes at $\xi=0$ (marked by different symbols) and display them in Fig.~\ref{fig:local}B. The high decay rates and localization at opposite edges confirm that these branches correspond to non-polarized edge modes. We then extend our analysis to a finite domain, where we compute the eigenfrequencies and classify the resulting modes as bulk, edge, or corner states using the color-coded scheme shown in Fig.~\ref{fig:local}C. We also highlight two representative mode shapes: an edge mode within the first BG and a corner mode within the second BG.

\section{Effective Elastic Moduli of the Optimized Designs}
From the acoustic modes in the band diagrams of each optimized configuration, we estimate the corresponding effective elastic moduli~\cite{phani2006wave}. The effective mass density is given by $\rho^*=\rho V$, where $\rho$ is the mass density of aluminum and $V$ is the volume fraction, obtained by dividing the solid area by the total area of the hexagonal design space. Next, we numerically determine the pressure and shear wave velocities, $(c_p,c_s)$, which correspond to the asymptotic slopes of the dispersion curves in the long-wavelength limit $( \|\mathbf{k}\| \xrightarrow{} 0)$. The HSP $\Gamma$ represents this limit, where an effective medium approximation is valid. Two modes emerge from $\Gamma$, corresponding to longitudinal and transverse waves, with their slopes defining $c_p$ and $c_s$, respectively. Notably, the similar slopes along the $\Gamma\xrightarrow{}$M and $\Gamma\xrightarrow{}$K directions suggest near-isotropic behavior in this regime.

\begin{figure}[h!]
\includegraphics[width=0.9\columnwidth]{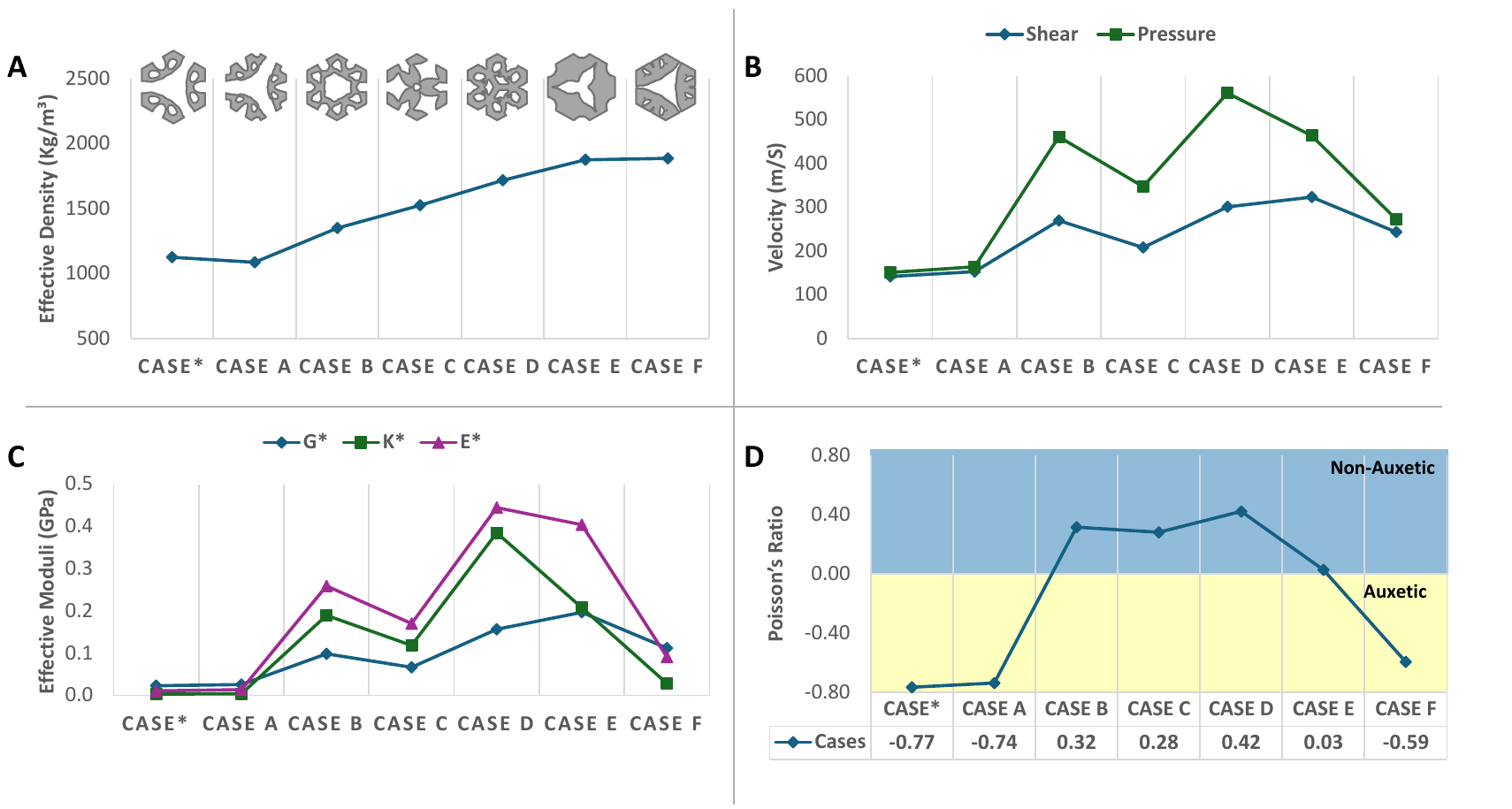}
\caption{\label{fig:mod}Effective elastic moduli of the optimized designs. (A) Effective mass density, (B) shear and pressure wave velocities, (C) effective shear $G^*$, bulk $K^*$, and Young’s moduli $E^*$, and (D) effective Poisson’s ratio $\nu^*$ for the different configurations.}
\end{figure}

The effective shear modulus $G^*$, bulk modulus $K^*$, and Young's modulus $E^*$ are computed from the wave velocities using the relations for a linear, homogeneous, and isotropic medium: $c_p=\sqrt{(K^*+G^*)/\rho^*}$, $c_s=\sqrt{G^*/\rho^*}$, and $E^*=4K^*G^*/(K^*+G^*)$. The effective Poisson’s ratio $\nu^*$ follows from the relation: $\nu^* = (K^* + G^*)/(K^*-G^*)$ for 2D plane stress. Fig.~\ref{fig:mod} presents the results for the main configuration studied throughout the manuscript (Case$^*$) and the library of designs introduced in Fig.~5 of the main article (Cases A–F). In particular,  Case$^*$, Case A, and Case F exhibit auxetic behavior, where Poisson's ration $\nu^*$ is negative and the bulk modulus $K^*$ approaches zero. Interestingly, these cases share structural similarities with kagome lattices, which are deeply auxetic $\nu=-1$ and exhibit equal pressure and shear wave velocities $(c_p=c_s)$, leading to a vanishing bulk modulus.

\end{document}